\documentclass[aps,preprintnumbers,superscriptaddress,showpacs]{revtex4}
\usepackage{epsfig}
\usepackage{psfrag}
\usepackage{amsfonts}
\usepackage{graphicx}
\usepackage{dcolumn}
\usepackage{bm}

\begin{document}

\title{Heavy quark potential from deformed $AdS_5$ models}

\author{Zi-qiang Zhang}
\email{zhangzq@cug.edu.cn} \affiliation{School of mathematics and
physics, China University of Geosciences(Wuhan), Wuhan 430074,
China}

\author{De-fu Hou}
\email{houdf@mail.ccnu.edu.cn} \affiliation{Key Laboratory of
Quark and Lepton Physics (MOE), Central China Normal University,
Wuhan 430079,China}

\author{Gang Chen}
\email{chengang1@cug.edu.cn} \affiliation{School of mathematics
and physics, China University of Geosciences(Wuhan), Wuhan 430074,
China}
\begin{abstract}
In this paper, we investigate the heavy quark potential in some
holographic QCD models. The calculation relies on a modified
renormalization scheme mentioned in a previous work of Albacete et
al. After studying the heavy quark potential in Pirner-Galow model
and Andreev-Zakharov model, we extend the discussion to a general
deformed $AdS_5$ case. It is shown that the obtained potential is
negative definite for all quark-antiquark separations, differs
from that using the usual renormalization scheme.
\end{abstract}


\maketitle
\section{Introduction}

Heavy quark potential is an important quantity in the study of
Quark Gluon Plasma(QGP) in heavy ion collisions carried out at
RHIC or LHC, since the melting of heavy mesons is usually
considered to be one of the signatures for QGP formation. However,
much experiment data indicates that QGP is strongly coupled
\cite{JA,KA,EV}. Thus, calculational tools for strong coupled,
real-time QCD dynamics are requried. Such tools are now available
via the AdS/CFT correspondence
\cite{Maldacena:1997re,Gubser:1998bc,MadalcenaReview}.

AdS/CFT, the duality between the type IIB superstring theory
formulated on AdS$_5\times S^5$ and $\mathcal N=4$ SYM in four
dimensions, has yielded many important insights into the dynamics
of strongly-coupled gauge theories. By using the AdS/CFT,
Maldacena has calculated the heavy quark potential in vacuum for
$\mathcal N=4$ SYM firstly \cite{Maldacena:1998im}. There, it is
found that for the $AdS_5$ space the energy shows a purely
Coulombian (non confining) behavior, agrees with a conformal gauge
theory. This proposal has attracted lots of interest. Soon after
\cite{Maldacena:1998im}, there are many attempts to addressing
heavy quark potential in this direction. For instance, the
potential at finite temperature has been discussed in
\cite{AB,SJ}. The potential in different spaces is investigated in
\cite{YK}. The subleading order corrections to this quantity are
considered in \cite{SX,ZQ}. This quantity is also analyzed with
some AdS/QCD models \cite{OA2,SH,DF,HJ}. Other important results
can be found, for example, in \cite{JG,FB,LM}.

Interestingly, the heavy quark potential can also be obtained by
using an alternative renormalization scheme \cite{AL}. There,
instead of subtracting the self energy of two static quarks, it
subtracts the real part of the action of the two quarks that
separated by an infinite distance. It was argued \cite{AL} that
for AdS/CFT the obtained potential of the two renormalization
schemes is the same at zero temperature but differs at finite
temperature. For finite temperature case, they find a smooth
non-zero (negative definite) potential without a kink, and the
potential has a non-zero imaginary (absorptive) part for
separations $r>r_c=0.870/{\pi T}$. This result is obviously
different from that using the usual renormalization procedure
\cite{AB,SJ}, where the resulting potential goes to zero above
$r>r_*=0.754/{\pi T}$.

As the alternative renormalization is very interesting and the
heavy quark potential based on it has been discussed in the case
of AdS/CFT \cite{AL}, it is also of interest to extend the studies
of \cite{AL} to the case of AdS/QCD. In this paper, we would like
to see whether we can get a new and interesting result by using
this alternative renormalization for AdS/QCD. In addition, we will
compare the results with those from the usual renormalization
scheme and explore the reasonability of the results by comparing
with the experimental data. These are the main motivations of the
present work.

The paper is organized as follows. In the next section, some
holographic QCD models are briefly reviewed. In section 3, using
the modified renormalization procedurewe, we calculate the heavy
quark potential in Pirner-Galow model, Andreev-Zakharov model and
a general deformed $AdS_5$ model one by one. The summary and
discussion is given in section 4.

\section{Deformed AdS Backgrounds}
In this section, we briefly review some deformed AdS backgrounds
in \cite{SH1}. There are two main approaches in searching for the
string description of realistic QCD: ¡±top-down¡± approach, i.e.
by deriving holographic QCD from string theory
\cite{JBJ,MKD,TSS,SH2}, and ¡±bottom-up¡± approach, i.e. by
examining holographic QCD from experimental data and lattice
results \cite{HJ,OA1,AKE,JPF,KGN,UGE}.

In the ¡±bottom-up¡± approach, the most economic way is to use a
deformed $AdS_5$ metric, which can describe the known experimental
data and lattice results. The simplest holographic QCD model is
the hard-wall $AdS_5$ model \cite{JEE}, which has a good
description of the lightest meson spectra comparing with the
experimental data, but cannot produce the Regge behavior for
higher excitations. The soft-wall model \cite{AKE} suggests that a
negative quadratic dilaton term $-z^2$ in the action can produce
the right linear Regge behavior of $\rho$ mesons or the linear
confinement. To produce linear behavior of heavy flavor potential,
a positive quadratic term modification \cite{OA1} is suggested to
be added in the deformed warp factor of the metric in
Andreev-Zakharov model \cite{OA3}. To consider the QCD running
coupling effect into the modified metric, see in \cite{UGE,HJ}. To
take into account the QCD $\beta$ function, see in \cite{SH1}.

\section{holographic heavy quark potential}
Before clarifying the definition of the heavy quark potential, we
briefly introduce the Polyakov loop for $SU(N_c)$ gauge theory. At
the spatial location $\vec{r}$, the Polyakov loop is
\begin{equation}
L(\vec{r})=\frac{1}{N_c}Tr[Pexp(ig\int_0^\beta d\tau
A_4(\vec{r},\tau))],
\end{equation}
where $\tau$ is the Euclidean time, T is the temperature and
$\beta=1/T$.

The singlet $V_1(r)$ and adjoint $V_{adj}(r)$ in Euclidean space
can de extracted from the connected correlator of two Polyakov
loops \cite{LD,SN}
\begin{equation}
<L(0)L^\dag(\vec{r})>_c=\frac{e^{-\beta V_1(r)}+(N_c^2-1)e^{-\beta
V_{adj}(r)}}{N_c^2}.
\end{equation}

To obtain the Polyakov loop correlator in AdS space, one needs to
connect the open strings to the positions of Polyakov loops at the
boundary of the AdS space in all possible ways
\cite{Maldacena:1998im,AB,SJ}. There are two relevant
configurations, hanging string and straight string, which can give
two different saddle points of the Nambu-Goto action,
$S_{NG}^{hanging}$ and $S_{NG}^{straight}$. In the large $N_c$ and
large $\lambda$ limit, we have
\begin{equation}
<L(0)L^\dag(\vec{r})>_c\propto\frac{e^{-S_{NG}^{hanging}}+(N_c^2-1)e^{-S_{NG}^{straight}}}{N_c^2}.
\end{equation}

Note that $S_{NG}^{straight}$ gives $V_{adj}(r)$ and
$S_{NG}^{hanging}$ gives $V_1(r)$. In lattice simulations,
$V_1(r)$ is regarded as the heavy-quark potential at finite
temperature \cite{CG}.

On the other hand, $V_1(r)$ can be extracted from the expectation
value of the static Wilson loop operator between a static pair of
$Q\bar{Q}$
\begin{equation}
e^{-i\mathcal {T}V_1(r)}\sim \lim_{\mathcal {T}\to \infty} <W(C)>,
\end{equation}
with
\begin{equation}
W(C)=\frac{1}{N}TrPe^{i\oint_C dx^\mu A_\mu},
\end{equation}
where C is a closed loop in a 4-dimensional spacetime, which is
usually taken as a rectangular loop of spatial and temporal
extensions r and $\mathcal {T}$, respectively. $A_\mu$ is the
non-Abelian gauge potential operator. $P$ enforces the path
ordering along the loop $C$.

As a result, the heavy quark potential $V_1(r)$ (henceforth we
call it $V_{Q\bar{Q}})$ can be written as
\begin{equation}
V_{Q\bar{Q}}=\frac{S_c}{\mathcal {T}}\label{v}.
\end{equation}
where $S_c$ is a regularized action associated with
$S_{NG}^{hanging}$.

\subsection{In Pirner-Galow model}

The background metric of the Pirner-Galow model can be written as
\cite{HJ}:
\begin{equation}
ds^2=h(z)\frac{R^2}{z^2}(-dt^2+d\vec{x}^2+dz^2),\label{metric}
\end{equation}
where
\begin{equation}
h(z)=\frac{log(\frac{1}{\epsilon})}{log[\frac{1}{(\Lambda
z)^2+\epsilon}]},\label{hz}
\end{equation}
with the IR sigularity at
\begin{equation}
z_{IR}=\sqrt{\frac{1-\epsilon}{\Lambda^2}},
\end{equation}
where $\Lambda$ is related to the $AdS_5$ radius as $\Lambda=1/R$,
$\epsilon$ denotes a dimensionless parameter, that is
\begin{equation}
\epsilon\equiv l_s^2\Lambda^2,
\end{equation}
the parameter $l_s$ can guarantee the conformal limit
$h(z)\rightarrow$ 1 at $z\rightarrow0$.

The Pirner-Galow model has some properties: (1). The functional
form of the warping factor (\ref{hz}) can coincide with the
functional form of the QCD running coupling. (2). The new
parameter $\epsilon$ can be fitted by reproducing the short and
long range Cornell potential. Due to this characteristic,
evaluation of the heavy quark potential in this background can be
considered as a good test of AdS/QCD.

We now study the heavy quark potential in Pirner-Galow model using
the metric of (\ref{metric}). The string action can reduce to the
Nambu-Goto action
\begin{equation}
S=-\frac{1}{2\pi\alpha^\prime}\int d\tau
d\sigma\sqrt{-detg_{\alpha\beta}}, \label{NG}
\end{equation}
with
\begin{equation}
g_{\alpha\beta}=G_{\mu\nu}\frac{\partial
X^\mu}{\partial\sigma^\alpha} \frac{\partial
X^\nu}{\partial\sigma^\beta},
\end{equation}
where $X^\mu$ and $G_{\mu\nu}$ are the target space coordinates
and the metric respectively, and $\sigma^\alpha$ with $\alpha=0,1$
parameterize the world sheet. The string tension
$\frac{1}{2\pi\alpha^\prime}$ is related to the 't Hooft coupling
constant by
\begin{equation}
\frac{R^2}{\alpha^\prime}=\sqrt{\lambda}.
\end{equation}

By using the static gauge
\begin{equation}
t=\tau, \qquad \sigma=x,
\end{equation}
and supposing the radical for the radial direction
\begin{equation}
z=z(x),
\end{equation}
the metric (\ref{metric}) can reduce to the following form:
\begin{equation}
ds^2=\frac{R^2}{z^2}h(z)[-d\tau^2+(1+\dot{z}^2)dx^2]. \label{ac}
\end{equation}

Then the Euclidean version of Nambu-Goto action in (\ref{NG}) is
given by
\begin{equation}
S=\frac{\mathcal {T}R^2}{2\pi\alpha^\prime}\int dx
\frac{h(z)}{z^2}\sqrt{1+\dot{z}^2},
\end{equation}
where $\dot{z}=\frac{dz}{dx}$. Now we identify the Lagrangian as
\begin{equation}
\mathcal L=\frac{h(z)}{z^2}\sqrt{1+\dot{z}^2}.
\end{equation}

Note that $\mathcal L$ does not depend on x explicitly, so we have
the conserved quantity,
\begin{equation}
\mathcal L-\frac{\partial\mathcal L}{\partial\dot{z}}\dot{z}.
\end{equation}

The boundary condition at $x=0$ is:
\begin{equation}
\dot{z}=0, \qquad z=z_0,
\end{equation}
which yields
\begin{equation}
\frac{h(z)}{z^2\sqrt{1+\dot{z}^2}}=\frac{h(z_0)}{z_0^2},
\end{equation}
then a differential equation is derived,
\begin{equation}
\dot{z}=\frac{dz}{dx}=\sqrt{(\frac{h(z)z_0^2}{h(z_0)z^2})^2-1}\label{dotr}.
\end{equation}

By integrating (\ref{dotr}) the separation length $L$ of the
inter-quark is given by
\begin{equation}
L=2\int_0^{z_0}\frac{dz}{\sqrt{(\frac{h(z)z_0^2}{h(z_0)z^2})^2-1}}.\label{x}
\end{equation}

Plugging (\ref{dotr}) into (\ref{NG}), the action for the heavy
quark pair is obtained
\begin{equation}
S(L,T)=\frac{\mathcal
{T}R^2}{\pi\alpha^\prime}\int_0^{z_0}dz\frac{h(z)\sqrt{1+\dot{z}^2}}{z^2\dot{z}}.
\label{NG1}
\end{equation}

Usually, the regularization action is related to the subtraction
of the self energy of the two quarks, but we now use the following
subtraction \cite{AL}
\begin{equation}
V(L)=\frac{S(L,T)-Re[S(L=\infty,T)]}{\mathcal {T}}. \label{V}
\end{equation}

To compare with the case in \cite{HJ}, we choose $\epsilon=0.48$,
$\Lambda=0.264GeV$. It was argued \cite{HJ} that
$z_r\rightarrow1.85GeV^{-1}$ when $L\rightarrow\infty$. In other
words, the potential contains no imaginary part in the range
$0<z_0<z_r$. Then, from (\ref{NG1}) and (\ref{V}), the heavy quark
potential in Pirner-Galow model can be written as
\begin{eqnarray}
V=\frac{R^2}{\pi\alpha^\prime}\int_0^{z_0}dz\frac{h(z)\sqrt{1+\dot{z}^2}}{z^2\dot{z}}-\frac{R^2}{\pi\alpha^\prime}\int_0^{z_r}dz\frac{h(z)\sqrt{1+\dot{z_1}^2}}{z^2\dot{z_1}},\qquad
0\leq z_0\leq z_r,\label{V1}
\end{eqnarray}
where
\begin{equation}
\dot{z_1}=\sqrt{(\frac{h(z)z_r^2}{h(z_r)z^2})^2-1}.
\end{equation}

\begin{figure}
\centering
\includegraphics[width=8cm]{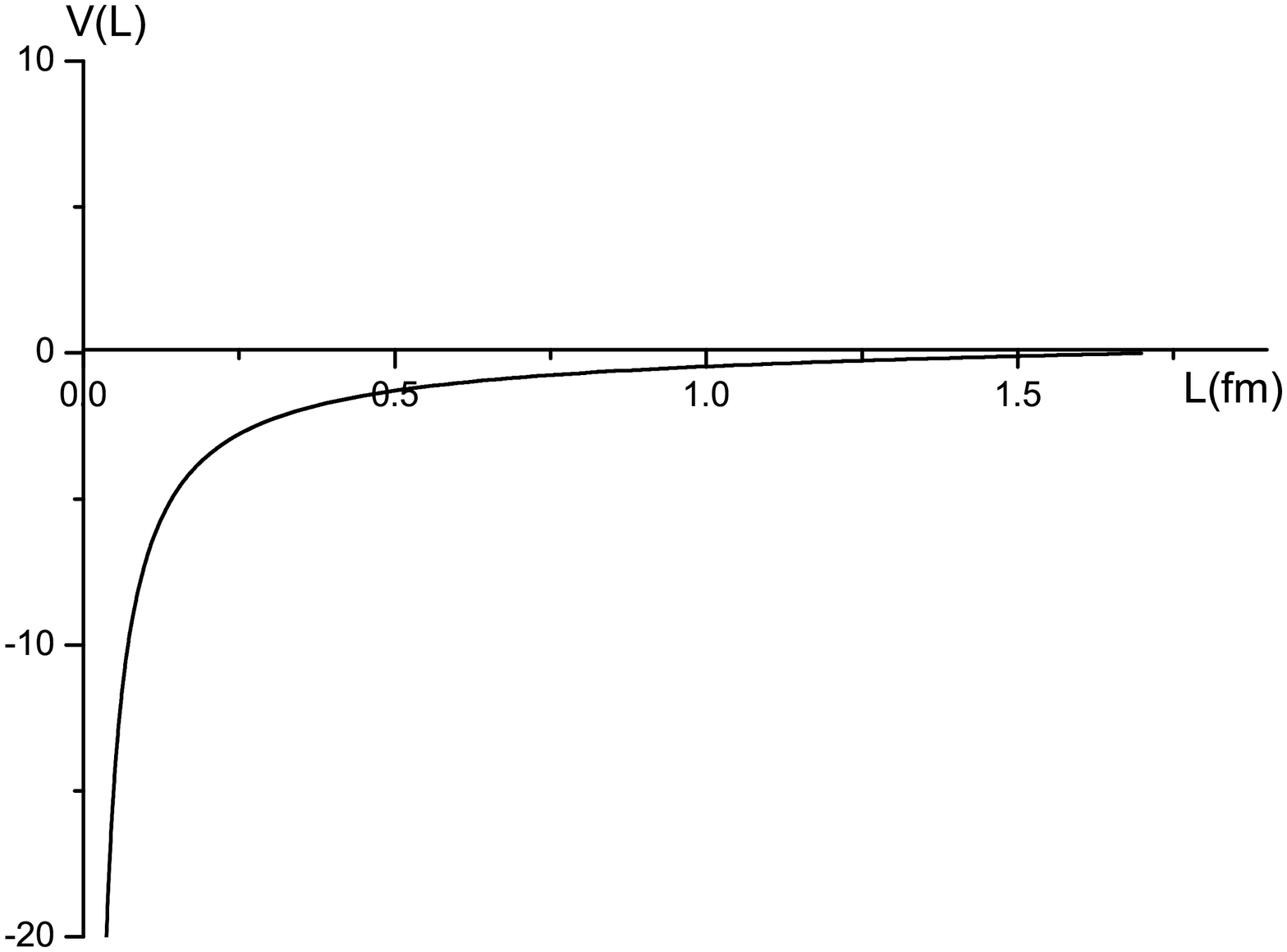}
\caption{Typical graph of $V(L)$. Here $\epsilon=0.48$,
$\Lambda=0.264GeV$.}
\end{figure}

To proceed further, we need to turn to numerical methods. Here we
plot the heavy quark potential V(L) against the inter distance L
in Fig.1. From the figure we can see clearly that the potential is
negative for all quark-antiquark separations, going to zero at an
infinite distance. If one uses the usual renormalization
procedure, the resulting potential goes to zero above a certain
critical separation and also has a kink \cite{HJ}.

\subsection{In Andreev-Zakharov model}

Next, we consider the heavy quark potential in Andreev-Zakharov
model. The corresponding background metric is given by \cite{OA3}
\begin{equation}
ds^2=\frac{R^2}{z^2}h(z)(-f(z)dt^2+d\vec{x}^2+\frac{1}{f(z)}dz^2),\label{metric1}
\end{equation}
with
\begin{equation}
h(z)=e^{\frac{cz^2}{2}},\qquad f(z)=1-\frac{z^4}{z_h^4},
\end{equation}
where $c$ is a deformation parameter. The horizon is located at
$z=z_h$ with $z_0<z_h$, where $z_0$ is a turning point or deepest
position of the string in the bulk. The temperature of the black
hole is given by
\begin{equation}
T=\frac{1}{\pi z_h}.
\end{equation}

The Andreev-Zakharov model has some properties: (1). In this
background linearized five-dimensional Yang-Mills equations are
effectively reduced to Laguerre differential equation, which can
help to fix the value of c from the $\rho$ meson trajectory
\cite{OA1}, therefore the metric contains no free parameter. (2).
The positive quadratic term modification in the deformed warp
factor $h(z)$ can produce linear behavior of heavy flavor
potential from the usual renormalization scheme.

We now explore the heavy quark potential in this metric
(\ref{metric1}) with the alternative renormalization scheme.
Parallel to the case of the previous Pirner-Galow model, we call
again the distance between quark anti-quark pair and the action
for the quark pair as $L$ and $S(L,T)$, respectively. One finds

\begin{equation}
L=2\int_0^{z_0}\frac{dz}{\sqrt{\frac{z_0^4}{z^4}e^{c(z^2-z_0^2)}\frac{(1-\frac{z^4}{z_h^4})^2}{(1-\frac{z_0^4}{z_h^4})}-(1-\frac{z^4}{z_h^4})}},\label{x1}
\end{equation}
and
\begin{equation}
S(L,T)=\frac{\mathcal
{T}R^2}{\pi\alpha^\prime}\int_0^{z_0}dz\frac{h(z)\sqrt{f+\dot{z}^2}}{z^2\dot{z}}.
\label{NG2}
\end{equation}

Likewise, $z_0$ also has a upper bound $z_{max}$, and the
potential does not contain an imaginary part. It has been verified
in \cite{OA2} that $z_{max}$ can be related to $z_h$ as
\begin{equation}
z_{max}=z_h\sqrt{\frac{2}{\sqrt{3}}sin(\frac{1}{3}arcsin\frac{T^2}{T_1^2})},
\end{equation}
with
\begin{equation}
T_1=\frac{1}{\pi}\sqrt{\frac{c}{\sqrt{27}}}.
\end{equation}

Then, from (\ref{V}) and (\ref{NG2}) the heavy quark potential in
Andreev-Zakharov model can be written as
\begin{eqnarray}
V=\frac{R^2}{\pi\alpha^\prime}\int_0^{z_0}dz\frac{h(z)\sqrt{f+\dot{z}^2}}{z^2\dot{z}}-\frac{R^2}{\pi\alpha^\prime}\int_0^{z_{max}}dz\frac{h(z)\sqrt{f+\dot{z_1}^2}}{z^2\dot{z_1}},\qquad
0\leq z_0\leq z_{max},\label{V2}
\end{eqnarray}
where
\begin{equation}
\dot{z_1}=\sqrt{\frac{z_{max}^4}{z^4}e^{c(z^2-z_{max}^2)}\frac{(1-\frac{z^4}{z_h^4})^2}{(1-\frac{z_{max}^4}{z_h^4})}-(1-\frac{z^4}{z_h^4})}.
\end{equation}

\begin{figure}
\centering
\includegraphics[width=8cm]{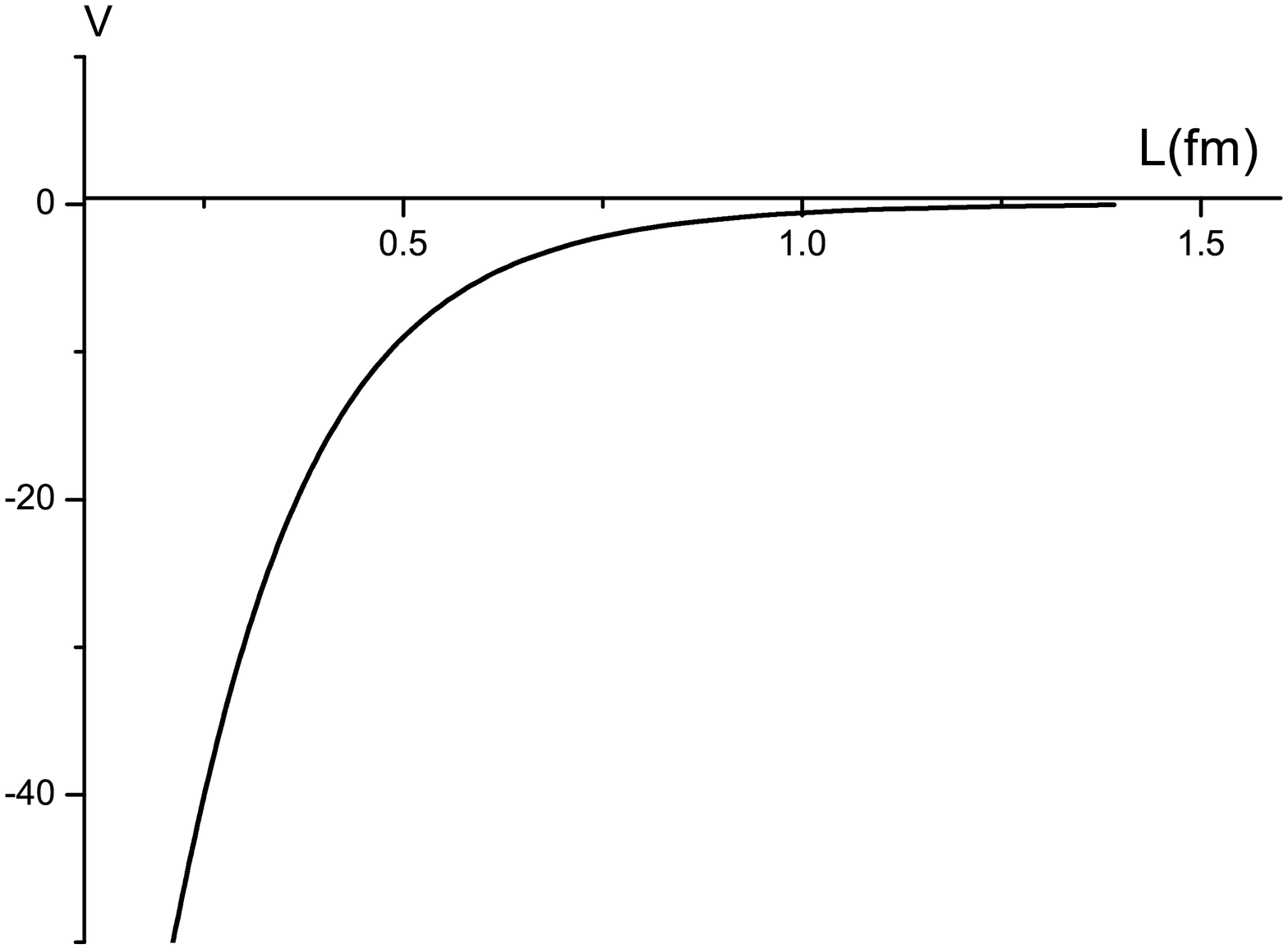}
\caption{Typical graph of $V$. Here $T=100Mev$, $c=0.9GeV^2$.}
\end{figure}

After taking $c=0.9GeV^2$ \cite{OA1}, we can plot the heavy quark
potential versus the separate distance of the quark anti-quark
numerically. From Fig.2 one finds the obtained potential is also
negative definite, differs from that using the usual
renormalization procedure \cite{OA2}.

\subsection{In a general deformed $AdS_5$ model}

To proceed, we extend the discussion to a general deformed $AdS_5$
model. The metric is given by
\begin{equation}
ds^2=\frac{R^2}{z^2}h(z)(-f(z)dt^2+d\vec{x}^2+\frac{1}{f(z)}dz^2),
\end{equation}
where the deformed warp factor $h(z)$ and the function $f(z)$
should satisfy the following conditions: (1). The conformal
invariance of QCD in UV limit requires that $h(0)=f(0)=1$. (2).
The existence of a horizon with a nonzero Hawking temperature
requires that $f(z)>0$ for $0<z<z_h$. With these restrictions, one
can get the action for the quark pair and the inter-quark distance
as
\begin{equation}
S(L,T)=\frac{\mathcal
{T}R^2}{\pi\alpha^\prime}\int_0^{z_0}dz\frac{h(z)\sqrt{f(z)+\dot{z}^2}}{z^2\dot{z}},\qquad
\dot{z}=\frac{dz}{dx}=\sqrt{(\frac{h(z)z_0^2}{h(z_0)z^2})^2\frac{f^2(z)}{f(z_0)}-f(z)},
\end{equation}
and
\begin{equation}
L=2\int_0^{z_0}\frac{dz}{\sqrt{(\frac{h(z)z_0^2}{h(z_0)z^2})^2\frac{f^2(z)}{f(z_0)}-f(z)}}.
\end{equation}

Likewise, the heavy quark potential can be found as
\begin{eqnarray}
V=\frac{R^2}{\pi\alpha^\prime}\int_0^{z_0}dz\frac{h(z)\sqrt{f+\dot{z}^2}}{z^2\dot{z}}-\frac{R^2}{\pi\alpha^\prime}\int_0^{z_{max}}dz\frac{h(z)\sqrt{f+\dot{z_1}^2}}{z^2\dot{z_1}},\qquad
0\leq z_0\leq z_{max},\label{V3}
\end{eqnarray}
where
\begin{equation}
\dot{z_1}=\frac{dz}{dx}=\sqrt{(\frac{h(z)z_{max}^2}{h(z_{max})z^2})^2\frac{f^2(z)}{f(z_{max})}-f(z)}.
\end{equation}

One can see that the potential of Eq.(\ref{V3}) must go to zero at
infinite separations (corresponding to $z_0=z_{max}$). Also, one
can check the results in some other holographic models such as the
Sakai-Sugimoto model \cite{TSS}, one may find that the result is
similar. Therefore, from this scheme the resulting potential seems
to be negative for all quark-antiquark separations. Of course, if
one doesn't use this specific ansatz, the result may not show this
behavior. Indeed, the resulting potential displays a kink-like
screening if one uses the usual renormalization procedure
\cite{YW}.

\section{summary and discussions}
In this paper, we have studied the heavy quark potential in some
deformed $AdS_5$ models using a modified renormalization
procedure. The quark-antiquark potential was obtained by
calculating the Nambu-Goto action of string attaching the
rectangular Wilson loop. It is found that the obtained potential
is negative for all quark-antiquark separations.

Before discussing the result, let us first recall the usual
renormalization procedure \cite{Maldacena:1998im}
\begin{equation}
V(L)=\frac{S(L,T)-2S_0}{\mathcal {T}}.\label{V0}
\end{equation}
where $S_0$ is the self energy or the mass of each quark. We can
see that the main difference between the two renormalization
procedures Eq.(\ref{V}) and Eq.(\ref{V0}) is the subtrahend in the
numerator.

Why the heavy quark potential obtained from the two
renormalization procedures is the same at zero temperature but
differs at finite temperature or in some deformed $AdS_5$ models?
One possible reason is that at zero temperature, the total energy
of the two quarks that separated by an infinite distance is
consistent with the self energy of the two quarks due to the
vacuum background. While in other backgrounds, the total energy of
the $Q\bar{Q}$ separated by an infinite distance may not only
contain the self energy but also contain some energy from the
field or the space such as the temperature, electric, etc.

Finally, we would like to discuss the reasonability of the
results. The obtained potential in these deformed $AdS_5$ models
does not level off, implying the $Q\bar{Q}$ pair never dissociates
in a thermal medium. The results seem to be in contradiction with
the experimental studies. However, in \cite{AL} the authors have
analytically continued the string configurations into the complex
plane. This manipulation yields a nonzero imaginary potential
which would be responsible for melting the $Q\bar{Q}$ \cite{JN}.
So if we do this in the case of AdS/QCD, we may also get the
imaginary potential. But due to the presence of the deformed warp
factors in deformed $AdS_5$ models, it is difficult to do this
manipulation analytically and one has to resort to numerical
methods. We hope to report our progress in this regard in future.

\section{Acknowledgments}

We would like to thank Prof Hai-cang Ren for useful discussions.
This research is partly supported by the Ministry of Science and
Technology of China (MSTC) under the ¡°973¡± Project no.
2015CB856904(4). Zi-qiang Zhang and Gang Chen are supported by the
NSFC under Grant no. 11475149. De-fu Hou is supported by the NSFC
under Grant no. 11375070, 11521064.


\end{document}